\documentclass[prd,amsmath,amssymb,superscriptaddress,onecolumn]{revtex4-2}

\usepackage{bbm}
\usepackage{todonotes}
\usepackage[export]{adjustbox}
\usepackage{graphicx}
\usepackage{epsfig}
\usepackage{braket}

\usepackage{bm}

\usepackage{capt-of}

\usepackage{hyperref}

\numberwithin{equation}{section}

\usepackage[noabbrev]{cleveref}

\newcommand{\bb}{\mbox{\boldmath $b$}}

\newcommand{\bba}{\mbox{\boldmath $b_{1}$}}
\newcommand{\bbb}{\mbox{\boldmath $b_{2}$}}

\newcommand{\bpp}{\mbox{\boldmath $p_{3}$}}
\newcommand{\bqq}{\mbox{\boldmath $q_{1}$}}

\begin{document}

\title{\boldmath 
Exclusive $D \bar{D}$ pair production with low invariant mass\\
in ultraperipheral Pb-Pb collisions at the LHC}

\author{Piotr Lebiedowicz}
\email{Piotr.Lebiedowicz@ifj.edu.pl}
\affiliation{Institute of Nuclear Physics Polish Academy of Sciences, Radzikowskiego 152, PL-31342 Krak{\'o}w, Poland}

\author{Antoni Szczurek}
\email{Antoni.Szczurek@ifj.edu.pl}
\affiliation{Institute of Nuclear Physics Polish Academy of Sciences, Radzikowskiego 152, PL-31342 Krak{\'o}w, Poland}
\affiliation{Institute of Physics, Faculty of Exact and Technical Sciences, University of Rzesz{\'o}w, 
Pigonia 1, PL-35310 Rzesz{\'o}w, Poland}

\begin{abstract}
We present predictions for the ultraperipheral heavy-ion reaction ${\rm Pb} {\rm Pb} \to {\rm Pb} {\rm Pb} D \bar{D}$, where $D$ refers to either $D^0$ or $D^+$, limiting to low invariant mass of $D \bar{D}$ and at energies available at the LHC. The calculation of the $\gamma \gamma \to D \bar{D}$ subprocess is done including the continuum mechanisms and the $\chi_{c0}(3860)$ and $\chi_{c2}(3930)$ resonant contributions. These states are considered as candidates for the first excited $P$-wave charmonia $\chi_{c0}(2P)$ and $\chi_{c2}(2P)$. We compare our results for the $\gamma \gamma \to D \bar{D}$ process with the $D \bar{D}$ invariant mass and angular distributions measured by the Belle and BaBar experiments in $e^{+}e^{-}$ collisions. Then, we present first calculations for ultraperipheral Pb-Pb collisions (UPCs) within the equivalent photon approximation in the impact-parameter space. Both, the total cross section and several differential distributions for experimental cuts corresponding to the ALICE, ATLAS, CMS, and LHCb experiments are presented. For instance, the nuclear cross sections for the $D^{0} \bar{D}^{0}$ and $D^{+} D^{-}$ channel at $\sqrt{s_{NN}} = 5.02$ TeV are found to be approximately 100--132 $\mu$b and 29 $\mu$b, respectively, taking into account the cuts on pseudorapidities and transverse momenta of the final state charmed mesons ($|\eta_{D}| < 2.5$, $p_{t,D} > 0.2$ GeV) and for $M_{D \bar{D}} < 4.3$ GeV. Results for $\sqrt{s_{NN}} = 5.36$ TeV are also presented. Cross section for the ${\rm Pb} {\rm Pb} \to {\rm Pb} {\rm Pb}  (\gamma \gamma \to \chi_{c0,2}(2P) \to D \bar{D})$ processes is sufficiently large for experimental studies. The back-to-back correlation between charm mesons can be used to separate the exclusive process under consideration from other background processes, and could provide valuable insight into the properties of the excited quarkonia.
\end{abstract}



\maketitle

\section{Introduction}
\label{sec:Introdution}

The exclusive production of $D \bar{D}$ meson pairs
in ultraperipheral heavy-ion collisions (UPCs),
$AA \to AA D \bar{D}$,
where $A$ refers to either ${\rm Au}$ or ${\rm Pb}$,
has been discussed previously in \cite{Luszczak:2011js}.
There, the elementary $\gamma \gamma \to D \bar{D}$ cross sections
were estimated in the heavy-quark approximation
\cite{Baek:1994kj}
and in the Brodsky-Lepage formalism 
\cite{Brodsky:1981rp}
with the $D$ meson distribution amplitude 
\cite{Zuo:2006re} describing CLEO data 
on leptonic $D^{+}$ decay.
Numerical predictions for the nuclear cross sections
were calculated in \cite{Luszczak:2011js}
in the equivalent photon approximation (EPA)
in the impact-parameter space,
in the kinematics case when the $D$-mesons have 
sufficiently large transverse momenta,
$p_{t,D} > 1$~GeV, and for large invariant masses
of the $D \bar{D}$ system, $M_{D \bar{D}} > 4$~GeV.
Rather small nuclear cross sections have been estimated,
for ${\rm Pb}$-${\rm Pb}$ collisions at LHC energy of $\sqrt{s_{NN}}=5.5$~TeV,
with the size of 0.28--4.09~$\mu$b for $D^0 {\bar D}^0$ 
and 1.68--1.92~$\mu$b for $D^+ D^-$
depending on the assumptions made in calculating elementary 
$\gamma \gamma \to D \bar{D}$ cross sections.
It was shown in Fig.~8 of \cite{Luszczak:2011js}
that the nuclear cross section 
$d\sigma/dM_{D \bar{D}}$
falling steeply with increasing $M_{D \bar{D}}$
with a size of order 1~$\mu$b/GeV for $M_{D \bar{D}} = 4$~GeV.
In the present Letter, 
we will discuss the production of $D \bar{D}$
in ultraperipheral heavy-ion collisions 
corresponding to $p_{t,D} \lesssim 1$~GeV
and at low invariant mass region.

Recently, in \cite{Babiarz:2025sld}
the reaction $e^+ e^- \to e^+ e^- (\gamma \gamma \to D \bar{D})$
was discussed,
along with a comparison of theoretical results
to the Belle \cite{Belle:2005rte}
and BaBar \cite{BaBar:2010jfn} data.
From this analysis it is clear that 
various mechanisms are needed to describe experimental results
measured in the $M_{D \bar{D}}$ region up to 4.3~GeV.
The first mechanism, called continuum contribution,
is described by $t$- and $u$-channel meson exchanges.
For the $D^0 {\bar D}^0$-continuum contribution,
the vector-meson $D^{*}(2007)^{0}$ exchanges were considered,
while in the $D^+ D^-$ case the $D^{\pm}$-meson exchanges 
were taken into account as the underlying continuum process.
The $D^{*}(2010)^{\pm}$ exchanges do not play an important role and can be safely neglected.
The second mechanism proceeds through 
an intermediate resonance states 
$\chi_{c0}(3860)$ and $\chi_{c2}(3930)$.
Here we assume that they correspond to 
the first excited $P$-wave charmonia, 
$\chi_{c0}(2P)$ and $\chi_{c2}(2P)$.
This theoretical approach describes
the $D \bar{D}$ invariant mass distributions 
for the $e^{+} e^{-} \to e^{+} e^{-} D \bar{D}$ reaction
including both the neutral and charged channels
measured by the BaBar Collaboration \cite{BaBar:2010jfn}. Comparison of the model results
to the efficiency-corrected (combined) BaBar data
is shown in Fig.~5 of \cite{Babiarz:2025sld}.
The $D^0 {\bar D}^0$-continuum contribution
plays an important role in describing the BaBar data.
One of the conclusions drawn in \cite{Babiarz:2025sld} 
is that the enhancement near the threshold is likely dominated 
by the continuum contribution 
rather than being attributed to the broad $\chi_{c0}(3860)$ resonance.
However, the interference effect between the $D^0 {\bar D}^0$ continuum 
and $\chi_{c0}(3860)$ contributions plays a significant role there.
This naturally raises the question of whether similar
situation occurs in the UPCs.
The possibility of identifying the $\chi_{c0}(2P)$ state, 
which is narrower than the $\chi_{c0}(3860)$ specified in PDG \cite{ParticleDataGroup:2024cfk},
and the $\chi_{c2}(2P)$ (a good candidate is $\chi_{c2}(3930)$)
also presents an interesting challenge for experimental studies.


The nature of the charmonium-like states, 
$\chi_{c0}(3860)$, $\chi_{c0}(3915)$, and $\chi_{c2}(3930)$, 
is not well understood;
see the discussion in \cite{ParticleDataGroup:2024cfk}
and the references listed in \cite{Babiarz:2025sld}.
The production of open-charm meson pairs in UPCs
is one of the potential opportunities to search 
for the exotic states containing charm;
see \cite{Sobrinho:2024tre} for some estimates in this matter.
In \cite{Sobrinho:2024tre}, the $D^{+}D^{-}$ production
was considered assuming $D \bar{D}$ molecule.
Further theoretical and experimental research is needed to verify the above assumptions for the $\chi_{c0,2}(2P)$ states and
to better understand their production and decay properties.

The Letter is organized as follows. 
In the next Section we discuss briefly
the theoretical framework for the $AA \to AA D \bar{D}$ reaction
and provide the amplitudes for the $\gamma \gamma \to D \bar{D}$ subprocess.
In Section~\ref{sec:Results} we estimate numerically
the nuclear cross section in the EPA approach.
Finally, in Section~\ref{sec:Conclusions} we draw conclusions.

\section{Theoretical formalism}
\label{sec:Formalism}

\begin{figure}
\includegraphics[width=0.3\textwidth]{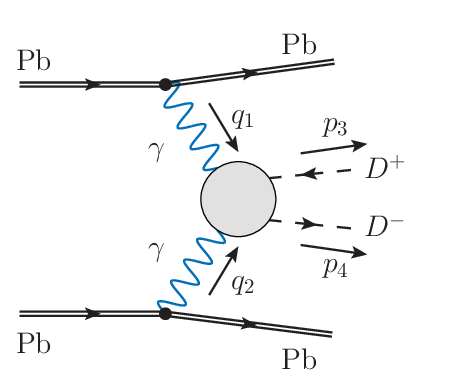}
\quad
\includegraphics[width=0.65\textwidth]{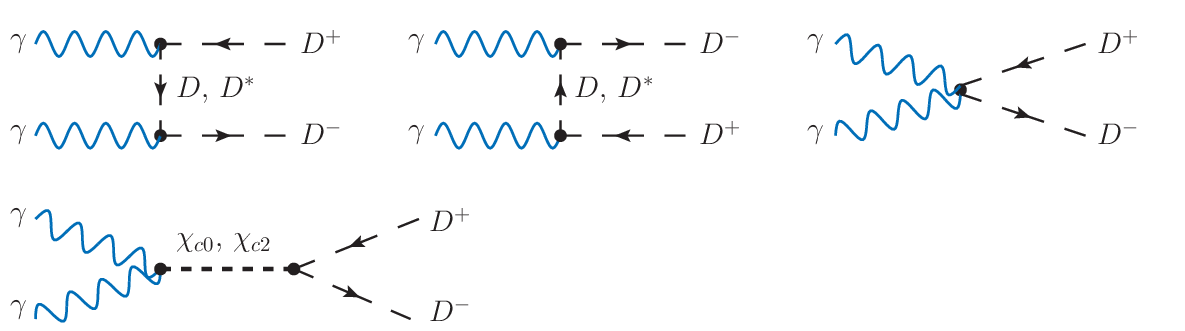}
\caption{Diagram representing $D^{+}D^{-}$ production 
in ultraperipheral $\rm Pb$-$\rm Pb$ collisions.
For the $\gamma \gamma \to D^{+} D^{-}$ process we consider 
the $t$- and $u$-channel $D^{\pm}$ and $D^{*}(2010)^{\pm}$ exchanges,
the contact term,
and the $s$-channel $\chi_{c0}$ and $\chi_{c2}$ exchanges.
The diagrams for the $D^0 {\bar D}^0$ production are similar, 
but in the case of continuum mechanism
we consider only the $D^{*}(2007)^{0}$ $t/u$-channel exchanges.}
\label{fig:diagram1}
\end{figure}
We focus on the $D \bar{D}$ pair production 
in ultraperipheral heavy-ion collisions,
see the diagram shown in Fig.~\ref{fig:diagram1}.
The nuclear cross section is calculated 
in the equivalent photon approximation (EPA) 
in the impact parameter space,
see e.g. \cite{Luszczak:2011js,KlusekGawenda:2010kx,Klusek-Gawenda:2016euz,Klusek-Gawenda:2017lgt}.

The total (phase-space integrated) nuclear cross section 
can be calculated as follows
[see (2.16) of \cite{KlusekGawenda:2010kx}]
\begin{eqnarray}
\sigma_{{\rm Pb Pb} \to {\rm Pb Pb} D\bar{D}}&=& 
\int \frac{d\sigma_{\gamma\gamma \to D\bar{D}}(W_{\gamma\gamma})}{d \cos\theta}  \,
N(\omega_{1}, b_{1}) N(\omega_{2}, b_{2}) \,
S^2_{\rm abs}(b)\,
\frac{W_{\gamma\gamma}}{2}\,
dW_{\gamma \gamma} \, 
d{\rm Y}_{D\bar{D}} \, 
d\overline{b}_x \, 
d\overline{b}_y 
\, 2 \pi \,b\,db \,
d \cos\theta\,,
\label{eq:sig_nucl_tot}
\end{eqnarray}  
where $b = |\bb| = |\bba-\bbb|$ is the impact parameter, 
i.e., the distance between colliding nuclei
in the plane perpendicular to their direction of motion.
The quantities 
$\overline{b}_x = (b_{1x}+b_{2x})/2$
and
$\overline{b}_y = (b_{1y}+b_{2y})/2$
are the components of the $\bba$ and $\bbb$ vectors 
which mark a point (distance from first and second nucleus) 
where photons collide and particles are produced.
$W_{\gamma\gamma} = \sqrt{4\omega_1\omega_2}$ 
is the invariant mass of the $\gamma\gamma$ system
($W_{\gamma\gamma} \equiv M_{D \bar{D}}$),
$\omega_{1} = \frac{W_{\gamma\gamma}}{2} 
\exp({\rm Y}_{D\bar{D}})$ and
$\omega_{2} = \frac{W_{\gamma\gamma}}{2} 
\exp(-{\rm Y}_{D\bar{D}})$
are the photon energies,
${\rm Y}_{D\bar{D}}
=\frac{1}{2}
\ln\frac{\omega_{1}}{\omega_{2}}
=\frac{1}{2}({\rm y}_{D} + {\rm y}_{\bar{D}})$
is the rapidity of the $D\bar{D}$ system
which is produced in the photon-photon collisions. 
Furthermore, $M_{D \bar{D}}^{2} = 
2 m_{\rm T}^{2} (1 + \cosh({\rm y}_{D}-{\rm y}_{\bar{D}}))$
with $m_{\rm T}^{2} = p_{t,D}^{2} + m_{D}^{2}$.
The photon flux for quasi-real photons
$N(\omega_{i}, b_{i})$, $i = 1, 2$,
attributed to each of the nuclei,
can be calculated in the equivalent photon approximation 
[see formula (45) of \cite{Hencken:1993cf},
but written in the convention of (2.18) of 
\cite{KlusekGawenda:2010kx}]:
%
\begin{eqnarray}
N(\omega, b) = \frac{Z^{2} \alpha_{\rm em}}{\pi^{2}} \frac{1}{\omega}
\left[ \frac{\omega}{\gamma_{\rm L}}
K_{1}\left(\frac{\omega b}{\gamma_{\rm L}} \right) - 
\sqrt{\frac{\omega^{2}}{\gamma_{\rm L}^{2}} + \Lambda^{2}}\,
K_{1}\left( 
b \sqrt{\frac{\omega^{2}}{\gamma_{\rm L}^{2}} + \Lambda^{2}}
\right)
\right]^{2}\,.
\label{eq:monopole_ff}
\end{eqnarray}  
Here, 
$\alpha_{\rm em}$ is the fine-structure constant,
$K_{1}(.)$ is the modified Bessel function of the second kind,
$\gamma_{\rm L} = \sqrt{s_{NN}}/(2 m_{N})$ is the nucleus Lorentz factor,
and $m_{N} = m_{A}/A$ the mass of a single nucleon.
We consider lead nuclei with charge $Z = 82$ 
and atomic number $A = 208$.
$\Lambda$ is a parameter in the parametrization of the monopole-type form-factor 
adjusted to reproduce the root-square radius of a nucleus.
For $\rm^{208}Pb$, we assume $\Lambda = 0.088$~GeV.
A more elaborate discussion 
on the nuclear form factor can be found in \cite{KlusekGawenda:2010kx}.
The absorption (survival) factor $S^{2}_{\rm abs}(b)$
ensures that only ultraperipheral collisions are considered,
i.e. the two colliding nuclei do not undergo nuclear breakup.
For first calculations it is sufficient to take 
the sharp cut-off model:
%
\begin{equation}
S^{2}_{\rm abs}(b) = \theta\left(b-(R_{1}+R_{2})\right),
\label{survival_factor}
\end{equation}
where the sum of the nuclear radii
$R_{1}+R_{2} \approx 14$~fm.
It should be noted that, 
for large $b$ (greater than twice the radius of the ions)
in which we are interested,
the results obtained using Eq.~(\ref{eq:monopole_ff})
are nearly identical to those obtained for extended charge distribution 
(i.e., with a realistic nuclear form factor, which is the Fourier transform of the charge distribution in the nucleus); 
see e.g. Figs.~7 and 9(b) of \cite{KlusekGawenda:2010kx}.

Let us now focus on elementary cross section 
for the reaction
\begin{eqnarray}
\gamma\,(q_{1},\epsilon_{1}) + \gamma\,(q_{2},\epsilon_{2}) \to D\,(p_{3}) + \bar{D}\,(p_{4}) \,,
\label{2to2}
\end{eqnarray}
where the four-momenta and the polarization vectors of the photons
are indicated in brackets.
In the following we adopt the model developed recently
in \cite{Babiarz:2025sld}.
The model takes into account two different continuum contributions depending on the channel under consideration, 
as well as the $\chi_{c0,2}(2P)$ resonance contributions, 
namely $\chi_{c0}(3860)$ and $\chi_{c2}(3930)$.
A reasonable fit was made to the Belle and BaBar data;
see \cite{Babiarz:2025sld}. 
This was done by adjusting certain model parameters.

The differential cross section for the reaction (\ref{2to2})
is given by
\begin{eqnarray}
&&\frac{d\sigma_{\gamma\gamma \to D\bar{D}}}{d \cos\theta} = \frac{1}{32 \pi W_{\gamma\gamma}^{2}} \frac{|\bpp|}{|\bqq|}
\frac{1}{4} \sum_{\rm spins}|{\cal M}_{\gamma\gamma \to D\bar{D}}|^{2}\,,
\label{2to2_xs}\\
&&
{\cal M}_{\gamma\gamma \to D\bar{D}}
= {\cal M}_{\mu \nu}^{(\gamma\gamma \to D\bar{D})} \epsilon_{1}^{\mu} \epsilon_{2}^{\nu}\,,
\end{eqnarray}
where 
$\theta$ denotes the angle of the outgoing meson
relative to the beam direction in the c.m. frame,
$\bqq$ and $\bpp$ are the c.m. three-momenta of the initial photon and final meson, respectively.
In the calculations
we consider the amplitudes ${\cal M}_{\mu \nu}$
for neutral and charged $D$-meson pair production
as a sum of the resonant and continuum processes:
\begin{eqnarray}
{\cal M}_{\mu \nu}^{(\gamma\gamma \to D^{0}\bar{D}^{0})}
&=& {\cal M}_{\mu \nu}^{(\chi_{c0}(3860) \to D^{0}\bar{D}^{0})}
+ {\cal M}_{\mu \nu}^{(\chi_{c2}(3930) \to D^{0}\bar{D}^{0})}
+ {\cal M}_{\mu \nu}^{(D^{0}\bar{D}^{0}\,{\rm continuum})}
\label{amp_gamgam_D0D0} \,,\\
{\cal M}_{\mu \nu}^{(\gamma\gamma \to D^{+}D^{-})}
&=& {\cal M}_{\mu \nu}^{(\chi_{c0}(3860) \to D^{+}D^{-})}
+ {\cal M}_{\mu \nu}^{(\chi_{c2}(3930) \to D^{+}D^{-})}
+ {\cal M}_{\mu \nu}^{(D^{+}D^{-}\,{\rm continuum})}\,.
\label{amp_gamgam_DpDm}
\end{eqnarray}
%

The amplitude for the $\gamma^* \gamma^* \to \chi_{c0}(3860) \to D \bar{D}$ 
process is given by
\begin{eqnarray}
i{\cal M}_{\mu \nu}^{(\chi_{c0}(3860) \to D \bar{D})}
=
i\Gamma^{(\gamma^* \gamma^* \to \chi_{c0})}_{\mu \nu}(q_{1},q_{2}) \,
i\Delta^{(\chi_{c0})}(p_{34})\,
i\Gamma^{(\chi_{c0} \to D \bar{D})}(p_{3},p_{4})\,,
\label{4.1}
\end{eqnarray}
where $p_{34} = q_{1} + q_{2} = p_{3} + p_{4}$, and
\begin{eqnarray}
i\Delta^{(\chi_{c0})}(p_{34})=
\frac{i}{p_{34}^{2}-M_{\chi_{c0}}^2+i M_{\chi_{c0}} \Gamma_{\chi_{c0}}}\,.
\label{4.1a}
\end{eqnarray}
We take
$M_{\chi_{c0}} = 3862~{\rm MeV}$ 
and $\Gamma_{\chi_{c0}} = 50~{\rm MeV}$ \cite{Babiarz:2025sld}.
The $\gamma^* \gamma^* \to \chi_{c0}$ vertex,
limiting only to the transverse component,
reads
\begin{equation}
\Gamma_{\mu \nu}^{(\gamma^* \gamma^* \to \chi_{c0})}(q_{1},q_{2}) =
-e^2 
R_{\mu \nu}
F_{TT}(Q_1^2,Q_2^2)\,,
\label{4.2}
\end{equation}
where 
$e^{2} = 4 \pi \alpha_{\rm em}$,
the photon virtualities
$Q_{1,2}^{2} \equiv -q_{1,2}^{2} \geqslant 0$, and
%
\begin{equation}
R_{\mu \nu} \equiv R_{\mu \nu}(q_1,q_2)
= -g_{\mu \nu} + \frac{1}{X}
\Big( (q_{1} \cdot q_{2}) (q_{1 \mu} q_{2 \nu} + q_{1 \nu} q_{2 \mu})
- q_{1}^{2} q_{2 \mu} q_{2 \nu} - q_{2}^{2} q_{1 \mu} q_{1 \nu} \Big)\,,
\label{2.2}
\end{equation}
$X = (q_{1} \cdot q_{2})^{2} - q_{1}^{2} q_{2}^{2}$. 
Since in UPC's $Q_{1}^2$, $Q_{2}^{2} \approx 0$ 
(large virtualities are suppressed by nuclear form factors) 
the terms $\propto q_{1}^2$, $q_{2}^{2}$ in (\ref{2.2}) can be safely neglected.
In the NRQCD approach, the form factor $F_{TT}$ 
in (\ref{4.2})
has the form
\begin{eqnarray}
F_{TT}(Q_1^2,Q_2^2) \to 
F_{TT}(0,0) = e_f^2 \sqrt{N_c} \frac{12}{\sqrt{\pi}}
\frac{R_{2P}'(0)}{M_{\chi_{c0}}^{3/2}}\,,
\label{4.4}
\end{eqnarray}
where $e_f = 2/3$ and $N_{c} = 3$.
The first derivative of the radial wave function at the origin is estimated for several $c \bar{c}$ interaction potential models from the literature,
see \cite{Babiarz:2025sld}.
For instance, for the Buchm\"uller-Tye potential model
one gets $|R'_{2P}(0)| = 0.326 ~ \rm GeV^{5/2}$,
while for the Cornell potential model one gets
$|R'_{2P}(0)| = 0.405 ~ \rm GeV^{5/2}$.
For the $\chi_{c0} D \bar{D}$ vertex we have ($M_{0} \equiv 1$~GeV)
\begin{eqnarray}
i\Gamma^{(\chi_{c0} \to D \bar{D})}(p_{3},p_{4})=
i g_{\chi_{c0} D\bar{D}} M_{0} \,
F^{(\chi_{c0})}(p_{34}^{2})\,.
\label{4.6}
\end{eqnarray}
In the calculation we put $F^{(\chi_{c0})}(p_{34}^{2}) = 1$.
The coupling constant $g_{\chi_{c0} D\bar{D}}$ is related to the partial
decay width of the $\chi_{c0}$ meson
\begin{eqnarray}
\Gamma{(\chi_{c0} \to D \bar{D})}=
\frac{M_{0}^{2}}{16 \pi M_{\chi_{c0}}} |g_{\chi_{c0} D\bar{D}}|^{2} 
\left( 1-\frac{4 m_{D}^{2}}{M_{\chi_{c0}}^{2}} \right)^{1/2}\,.
\label{4.7}
\end{eqnarray}
Taking
$\Gamma(\chi_{c0} \to D \bar{D}) = 16.4$~MeV
and using (\ref{4.7}) we get
$|g_{\chi_{c0} D\bar{D}}| = 2.52$ \cite{Babiarz:2025sld}.

The amplitude 
for the $\gamma^* \gamma^* \to \chi_{c2}(3930) \to D \bar{D}$ process 
is written as
\begin{eqnarray}
i{\cal M}_{\mu \nu}^{(\chi_{c2}(3930) \to D \bar{D})} 
&=& 
i\Gamma^{(\gamma^* \gamma^* \to \chi_{c2})}_{\mu \nu \rho \sigma}(q_{1},q_{2}) \;
i\Delta^{(\chi_{c2})\,\rho \sigma, \alpha \beta}(p_{34})\;
i\Gamma^{(\chi_{c2} \to D \bar{D})}_{\alpha \beta}(p_{3},p_{4})\,.
\label{3.1}
\end{eqnarray}
The propagator for the tensor $\chi_{c2}(3930)$ resonance is
\begin{eqnarray}
i\Delta_{\mu \nu, \kappa \lambda}^{(\chi_{c2})}(p_{34})&=&
\frac{i}{p_{34}^{2}-M_{\chi_{c2}}^2+i M_{\chi_{c2}} \Gamma_{\chi_{c2}}}
\left[ 
\frac{1}{2} 
( \hat{g}_{\mu \kappa} \hat{g}_{\nu \lambda}  + \hat{g}_{\mu \lambda} \hat{g}_{\nu \kappa} )
-\frac{1}{3} 
\hat{g}_{\mu \nu} \hat{g}_{\kappa \lambda}
\right] \,, 
\label{3.2}
\end{eqnarray}
where $\hat{g}_{\mu \nu} = -g_{\mu \nu} + p_{34 \mu} p_{34 \nu} / p_{34}^2$.
The total decay width of the $\chi_{c2}(3930)$ resonance and its mass is taken from the PDG~\cite{ParticleDataGroup:2024cfk}:
$M_{\chi_{c2}} = 3922.5 \pm 1.0~{\rm MeV}$ and
$\Gamma_{\chi_{c2}} = 35.2 \pm 2.2~{\rm MeV}$.
The $\gamma^* \gamma^* \to \chi_{c2}$ vertex,
limiting only to the helicity-2 component,
reads \cite{Babiarz:2025sld}
\begin{equation}
\Gamma_{\mu \nu \kappa \lambda}^{(\gamma^* \gamma^* \to \chi_{c2})}(q_{1},q_{2})=
e^2 
\frac{1}{2} \Big( R_{\mu \kappa}  R_{\nu \lambda}  
                + R_{\nu \kappa}  R_{\mu \lambda}
                - R_{\mu \nu}  R_{\kappa \lambda}  \Big) F_{TT,2}(Q_1^2,Q_2^2)\,,
\label{2.1}
\end{equation}
where, in the NRQCD limit and at $Q_{1,2}^2 = 0$, we have
%
\begin{eqnarray}
F_{TT,2}(Q_1^2,Q_2^2) \to 
%
%
F_{TT,2}(0,0) = 8 e_f^2 \sqrt{\frac{3 N_c}{\pi M_{\chi_{c2}}^{3}}} |R_{2P}'(0)|\,.
\label{2.12}
\end{eqnarray}
The $\chi_{c2} D \bar{D}$ vertex reads
($M_{0} \equiv 1$~GeV)
\begin{eqnarray}
i\Gamma_{\mu \nu}^{(\chi_{c2} \to D \bar{D})}(p_{3},p_{4})=
-i \,\frac{g_{\chi_{c2} D\bar{D}}}{2 M_{0}} 
\left[ (p_{3}-p_{4})_{\mu} (p_{3}-p_{4})_{\nu}
- \frac{1}{4} g_{\mu \nu} (p_{3}-p_{4})^{2} \right] F^{(\chi_{c2})}(p_{34}^{2})\,.
\label{3.4}
\end{eqnarray}
The value for the coupling constant
$g_{\chi_{c2} D\bar{D}} = 8.1 ~ (10.1)$
for the Cornell (Buchm\"uller-Tye) potential
is estimated by comparing model result
for the cross section
$\sigma(e^+ e^- \to e^+ e^- (\gamma^{*} \gamma^{*} \to \chi_{c2}(3930) \to D \bar{D}))$
to the corresponding BaBar result given by Eq.~(17) of \cite{BaBar:2010jfn};
see Section~3 of \cite{Babiarz:2025sld}.
Consequently, in the present calculations, 
we assume $F^{(\chi_{c2})}(p_{34}^{2}) = 1$.

The $\gamma \gamma \to D^{0} \bar{D}^{0}$ amplitude
with the $D^{*0} \equiv D^{*}(2007)^{0}$ 
$t/u$-channel exchanges can be expressed as
\begin{eqnarray}
i{\cal M}_{\mu \nu}^{(D^{0} \bar{D}^{0}\,{\rm continuum})}
&=& 
i\Gamma^{(D^{*0}D^0\gamma)}_{\kappa_{1} \mu}(\hat{p}_{t},q_{1})\,
i\tilde{\Delta}^{(D^{*0})\,\kappa_{1} \kappa_{2}}(p_{34}^{2},\hat{p}_{t}^{2})\,
i\Gamma^{(D^{*0}D^0\gamma)}_{\kappa_{2} \nu}(-\hat{p}_{t},q_{2})
\nonumber \\
&&
+ \,
i\Gamma^{(D^{*0}D^0\gamma)}_{\kappa_{1} \mu}(-\hat{p}_{u},q_{1})\,
i\tilde{\Delta}^{(D^{*0})\,\kappa_{1} \kappa_{2}}(p_{34}^{2},\hat{p}_{u}^{2})\,
i\Gamma^{(D^{*0}D^0\gamma)}_{\kappa_{2} \nu}(\hat{p}_{u},q_{2})\,,
\label{amplitude_D0D0bar_continuum}
\end{eqnarray}
where 
$\hat{p}_{t} = p_{a} - p_{1} - p_{3}$ and
$\hat{p}_{u} = p_{4} - p_{a} + p_{1}$.
The $D^{*0}D^{0}\gamma$ vertex, including a form factor, 
is taken as:
\begin{eqnarray}
i\Gamma_{\mu \nu}^{(D^{*0}D^0\gamma)}(\hat{p},q) 
= -i e \frac{g_{D^{*0}D^0\gamma}}{m_{D^{*0}}}\, 
\varepsilon_{\mu \nu \rho \sigma} \hat{p}^{\rho} q^{\sigma}
F^{(D^{*0}D^{0}\gamma)}(\hat{p}^{2},q^{2})\,.
\label{DstarDgam_vertex}
\end{eqnarray}
The coupling constant $|g_{D^{*0} D^{0} \gamma}| = 5.97$
is obtained from the decay width of $D^{*0}(2007)^{0} \to D^{0} \gamma$ assuming $\Gamma_{D^{*0}} = 55.3$~keV;
see \cite{Babiarz:2025sld}.
We use the factorized form for the $D^{*0}D^{0}\gamma$ form factor
\begin{eqnarray}
F^{(D^{*0}D^0\gamma)}(\hat{p}^{2},q^{2}) = 
F^{(D^{*0})}(\hat{p}^{2}) F^{(\gamma)}(q^{2}) 
\label{DstarDgam_ff}
\end{eqnarray}
with 
$F^{(D^{*0}D\gamma)}(m_{D^{*0}}^{2},0) = 1$.
In our case $q_{1,2}^{2} = 0$, and we take
\begin{eqnarray}
F^{(D^{*0})}(\hat{p}^{2}) &=& \exp \left( \frac{\hat{p}^{2}-m_{D^{*0}}^{2}}{\Lambda_{D^{*0}}^{2}} \right)\,,
\qquad \Lambda_{D^{*0}} = 3.3-3.5\;{\rm GeV}\,.
\label{DstarDgam_ff_exp}
\end{eqnarray}
In Eq.~(\ref{amplitude_D0D0bar_continuum}),
$\tilde{\Delta}^{(D^{*0})\, \kappa_{1} \kappa_{2}}(p_{34}^{2},\hat{p}_{t,u}^{2})$ denote
the propagators
of reggeized vector meson $D^{*}(2007)^{0}$.
For a detailed description of $\tilde{\Delta}^{(D^{*0})}$
see Eqs.~(43)--(46) of \cite{Babiarz:2025sld}.

For the $\gamma \gamma \to D^{+} D^{-}$ reaction,
as discussed in detail in \cite{Babiarz:2025sld},
the continuum mechanism proceeds via $D^{\pm}$ exchanges.
The corresponding amplitude reads
\begin{eqnarray}
i{\cal M}_{\mu \nu}^{(D^{+} D^{-}\,{\rm continuum})}
&=& -i e^{2}
\left[ 
(q_{1} - 2 p_{3})_{\mu}(q_{1} - p_{3} + p_{4})_{\nu} \frac{1}{\hat{p}_{t}^{2} - m_{D}^{2}}
+
(q_{1} - 2 p_{4})_{\mu}(q_{1} + p_{3} - p_{4})_{\nu} \frac{1}{\hat{p}_{u}^{2} - m_{D}^{2}}
-
2 g_{\mu \nu}
\right] \nonumber \\
&& \times 
F^{(\gamma)}(q_{1}^{2}) F^{(\gamma)}(q_{2}^{2})
F(\hat{p}_{t}^{2},\hat{p}_{u}^{2},p_{34}^{2})\,.
\label{amplitude_DpDm_continuum}
\end{eqnarray}
In the present case of UPC we have $F^{(\gamma)}(0) = 1$. 
The off-shell meson dependences 
are treated by a common form factor
$F(\hat{p}_{t}^{2},\hat{p}_{u}^{2},p_{34}^{2})$
[see (2.12), (2.13) of \cite{Klusek-Gawenda:2017lgt},
but with the replacements $m_{p} \to m_{D}$ and $\Lambda_{p} \to \Lambda_{D}$].
The parameter $\Lambda_{D}$ should be fitted to the experimental data.
It is assumed, however, that
$F(\hat{p}_{t}^{2},\hat{p}_{u}^{2},p_{34}^{2}) = 1$.

\section{Numerical results}
\label{sec:Results}

Before presenting the nuclear cross sections 
let us concentrate first on
elementary $\gamma \gamma \to D \bar{D}$ scattering.
In recent calculations 
we consider the (quasi)real photons
for the $\gamma \gamma \to D \bar{D}$ subprocess.
More specifically, in the calculation
of the elementary $\gamma \gamma \to D \bar{D}$ cross section here
the photons are treated as real 
(i.e., $Q_{1}^{2}, Q_{2}^{2} = 0$).

\begin{figure}
\includegraphics[width=0.325\textwidth]{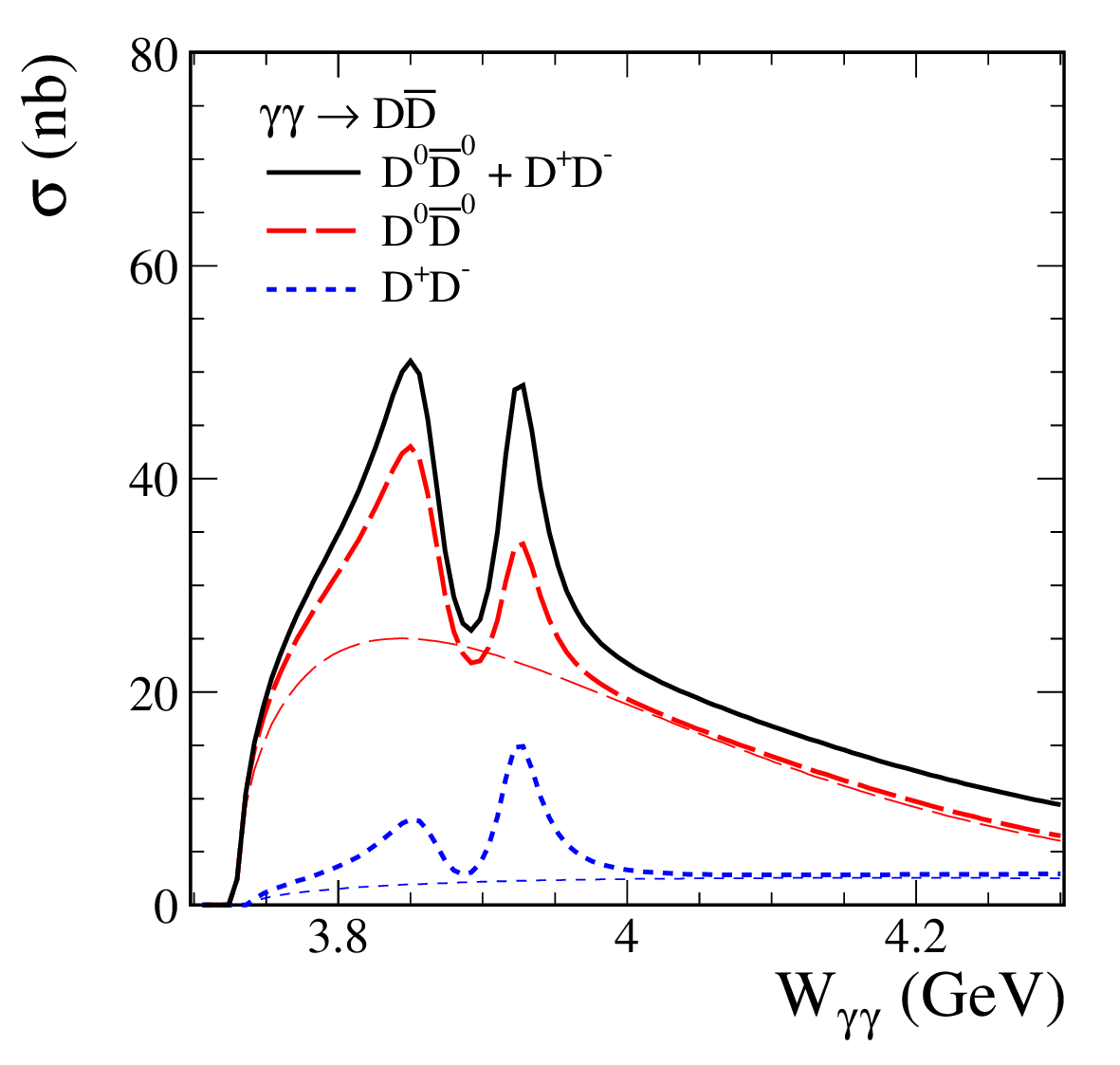}
\includegraphics[width=0.325\textwidth]{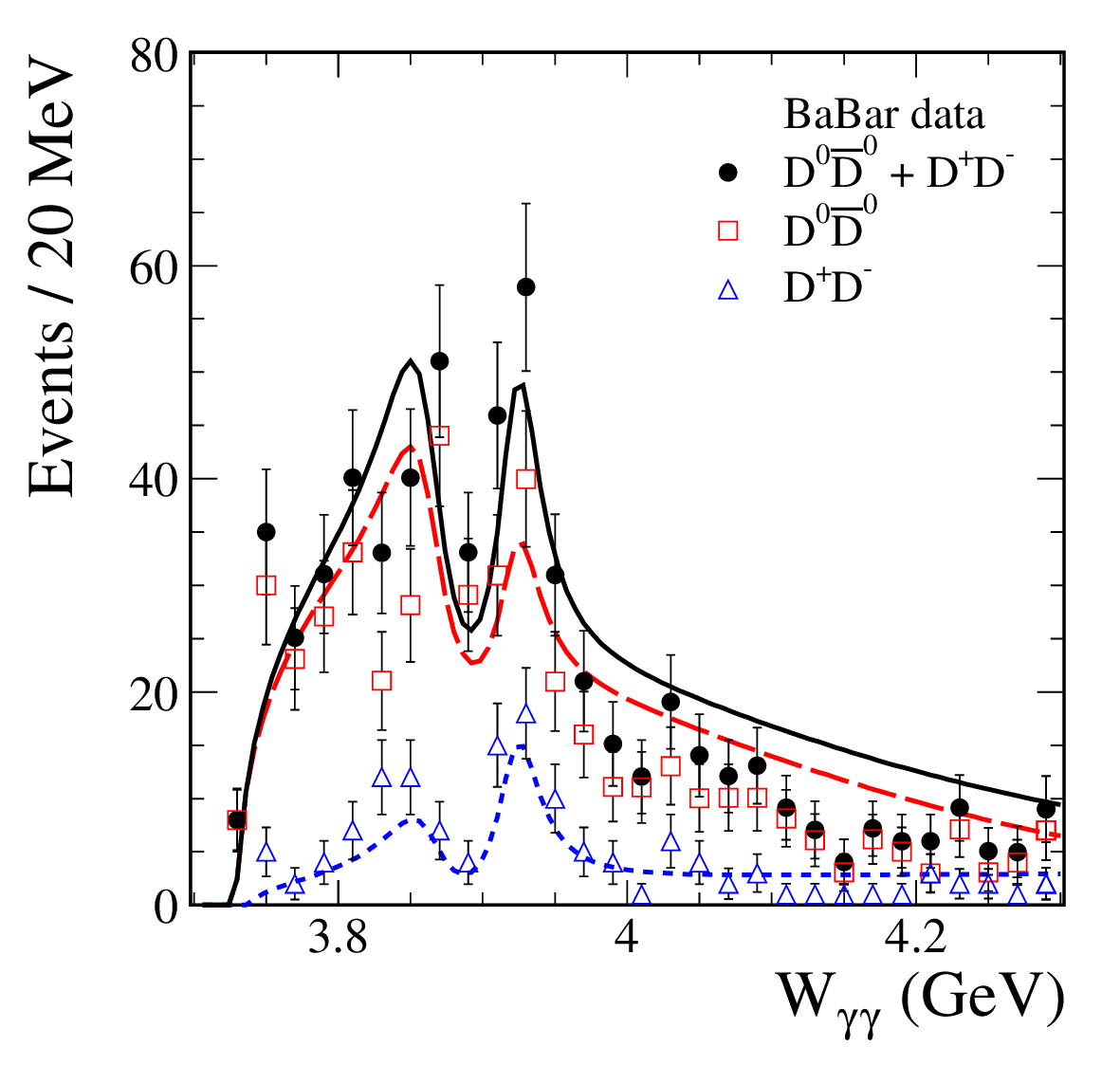}\\
\includegraphics[width=0.325\textwidth]{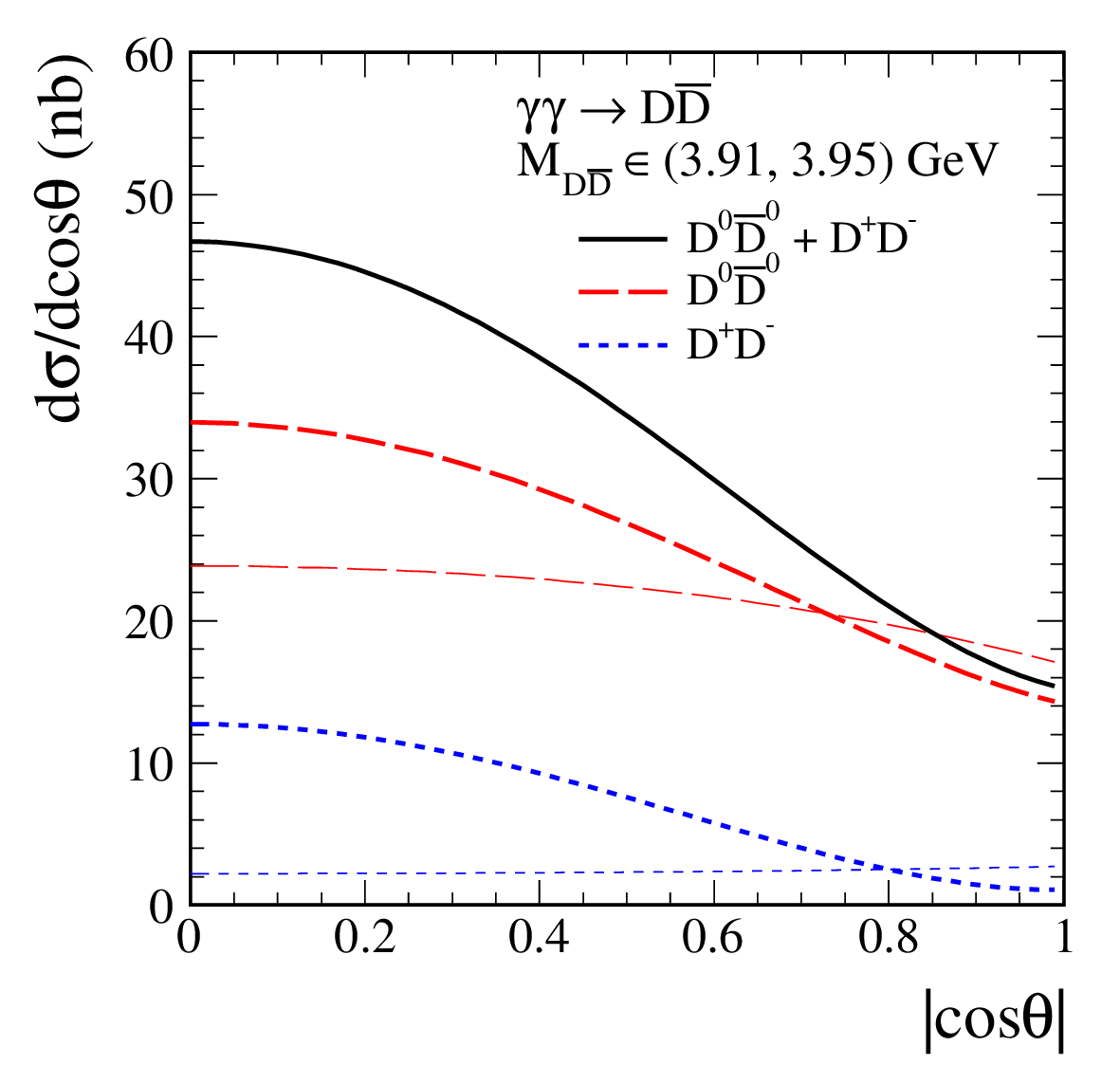}
\includegraphics[width=0.325\textwidth]{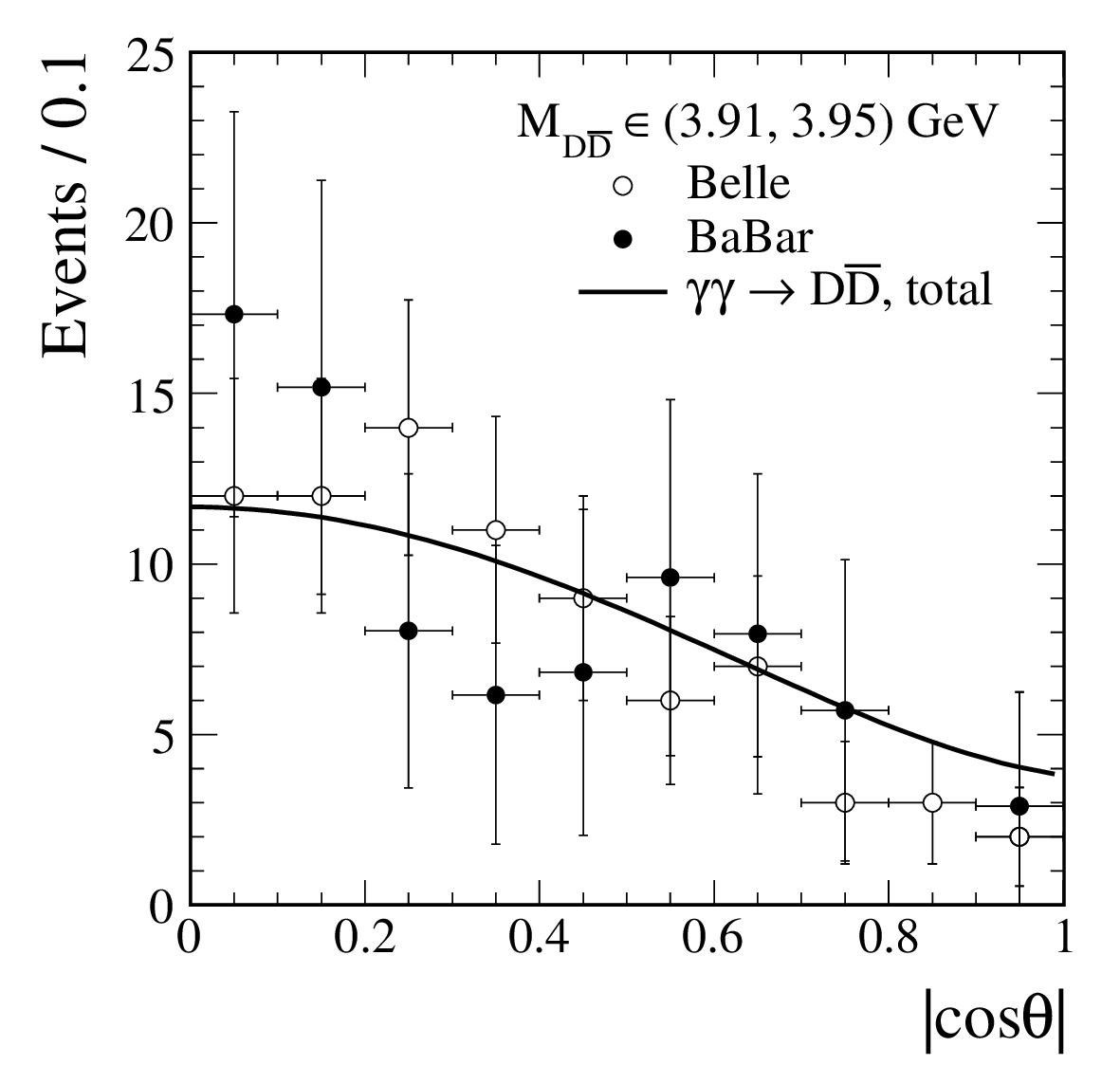}
\caption{The differential cross sections
for the $\gamma \gamma \to D \bar{D}$ processes
and comparison of theoretical results 
with the Belle \cite{Belle:2005rte}
and BaBar \cite{BaBar:2010jfn} data.
The theoretical results for the complete model 
(continuum, $\chi_{c0}(3860)$, and $\chi_{c2}(3930)$ contributions)
and for the continuum contribution alone are shown.
The red long-dashed lines correspond 
to the $D^0 {\bar D}^0$ channel
and the blue dashed lines correspond 
to the $D^{+}D^{-}$ channel.
The black solid line represents the sum of all contributions 
for the two channels.}
\label{fig:2to2}
\end{figure}

Figure~\ref{fig:2to2} shows 
the dependence of the $\gamma \gamma \to D \bar{D}$ 
cross sections on the photon-photon energy and
the angular distributions for 
$W_{\gamma \gamma} \in (3.91,3.95)$~GeV.
We show results for the $D^{0} \bar{D}^{0}$ 
and $D^{+} D^{-}$ channels separately,
and the combined $D \bar{D}$ final state
(denoted by $D^{0} \bar{D}^{0} + D^{+} D^{-}$).
We also compile the $\gamma \gamma \to D \bar{D}$ results 
with the Belle \cite{Belle:2005rte}
and BaBar \cite{BaBar:2010jfn} data
measured in $e^+ e^-$ collisions.
Comparing the above results for $\sigma(W_{\gamma \gamma})$
with the $d\sigma / dM_{D \bar{D}}$ cross sections
calculated for the $e^+ e^- \to e^+ e^- (\gamma^{*} \gamma^{*} \to D \bar{D})$ reactions \cite{Babiarz:2025sld}, 
we see that the former are above the latter, 
especially for larger values of 
$W_{\gamma \gamma}$ (or $M_{D \bar{D}}$).
As mentioned, above on-shell calculations are done 
without $F^{(\gamma)}(q_{1,2}^2)$ form factors 
and the terms $\propto q_{1,2}^2$ in the amplitudes.
Consequently, this leads to overestimation of the experimental data 
at larger $M_{D \bar{D}}$.
Therefore, our predictions for nuclear cross sections
for production of $D\bar{D}$ pairs, calculated in the EPA approach,
should be treated as upper estimates.

\begin{figure}
\includegraphics[width=0.325\textwidth]{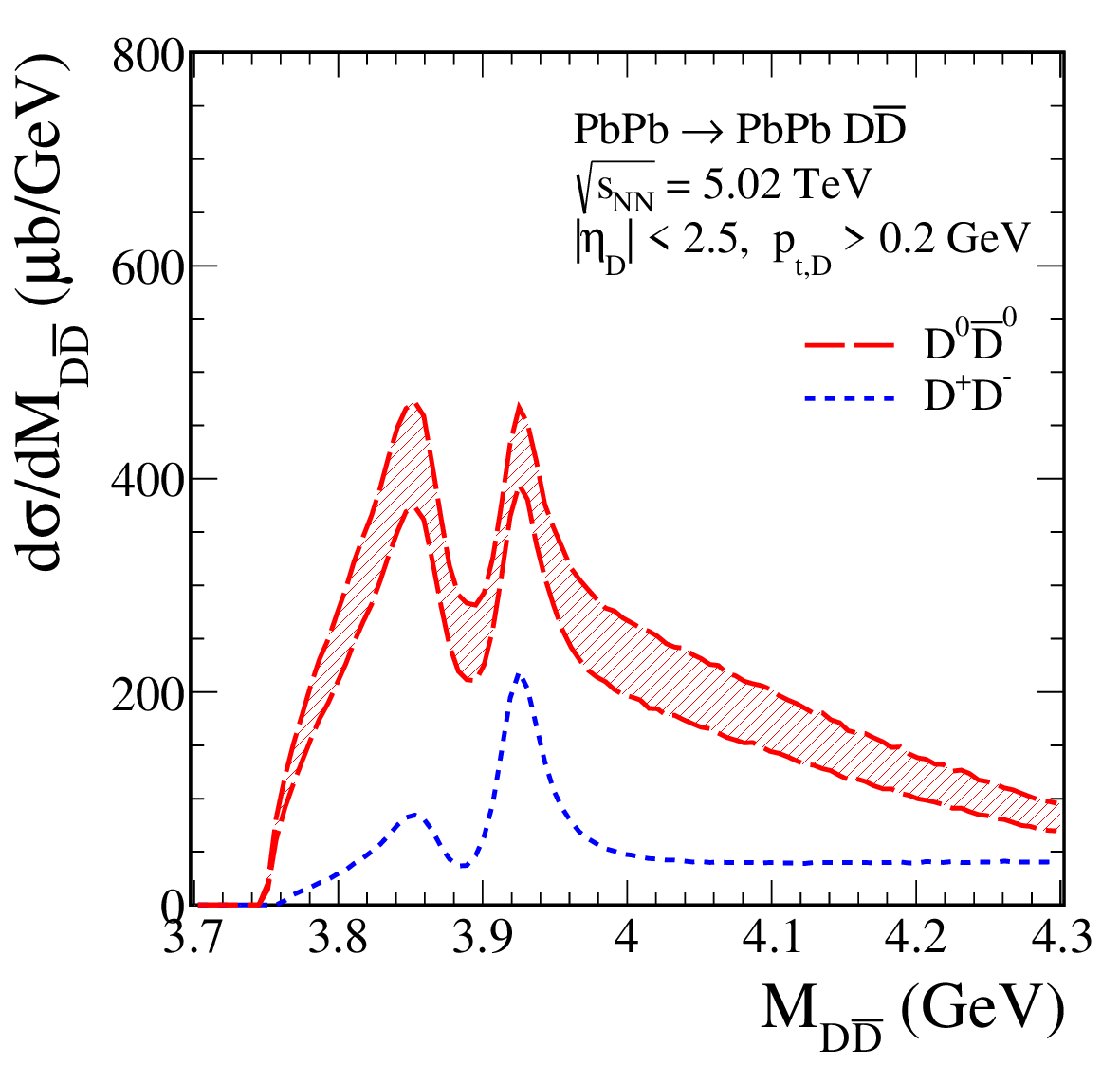}
\includegraphics[width=0.325\textwidth]{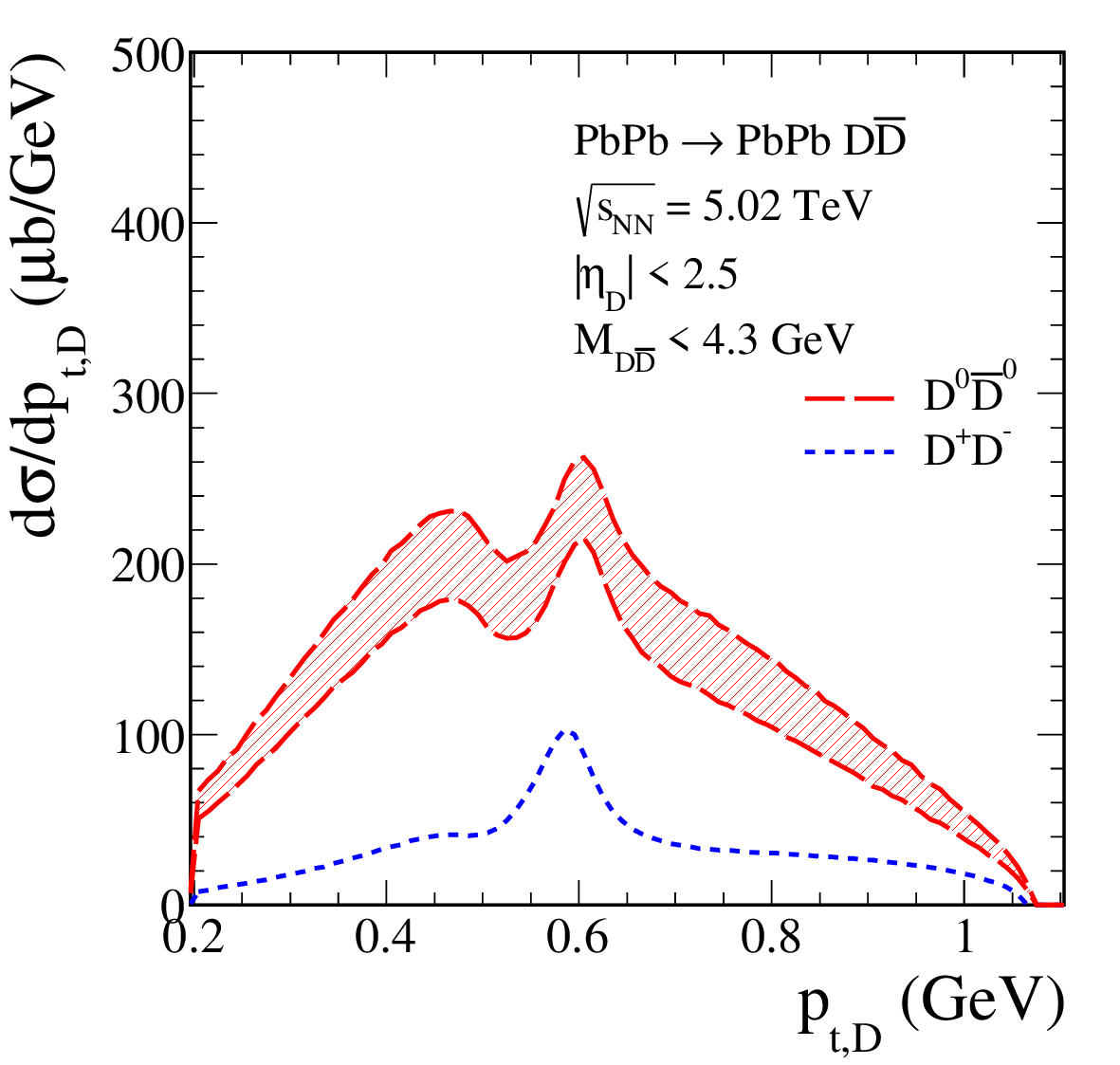}\\
\includegraphics[width=0.325\textwidth]{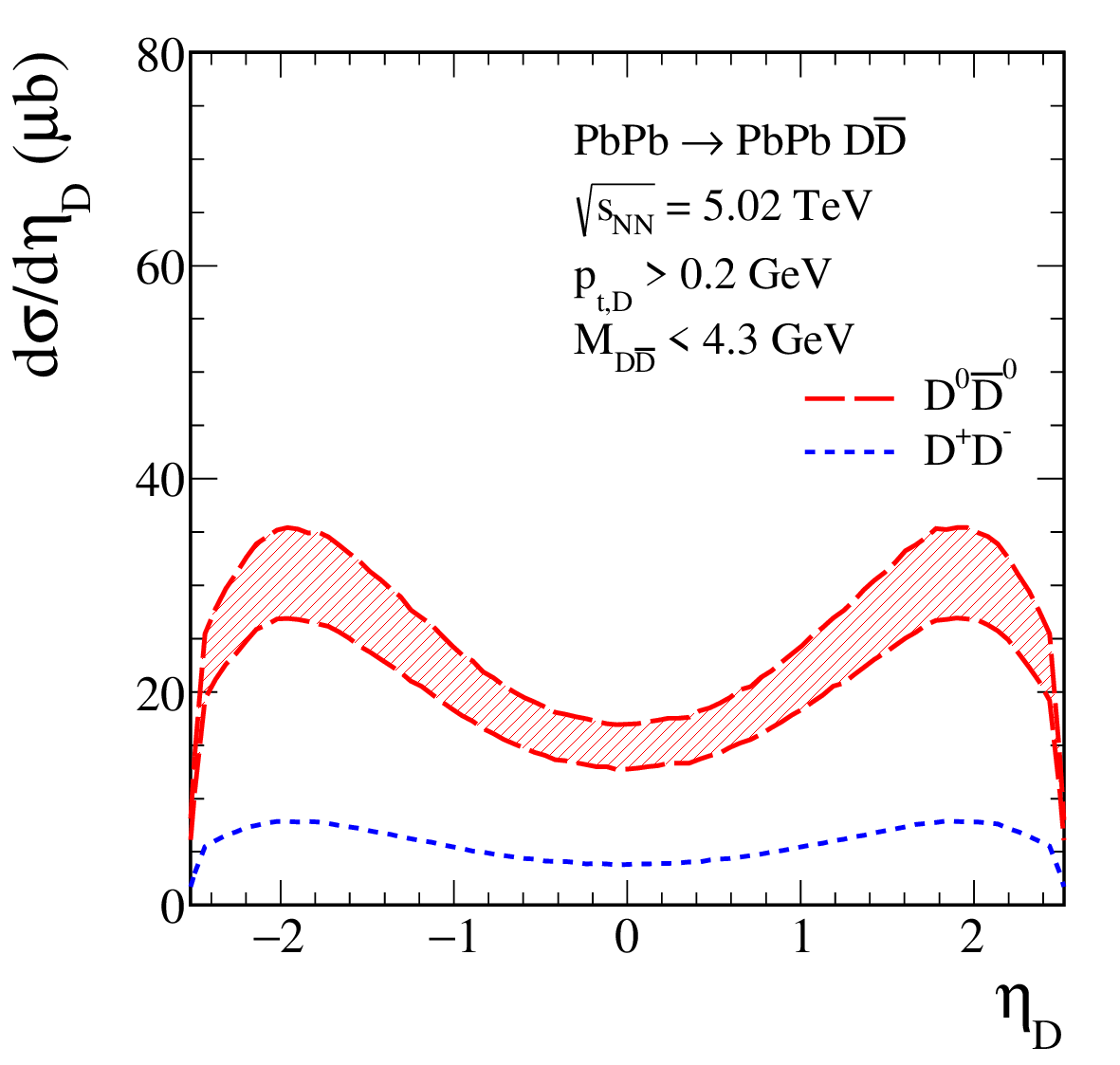}
\includegraphics[width=0.325\textwidth]{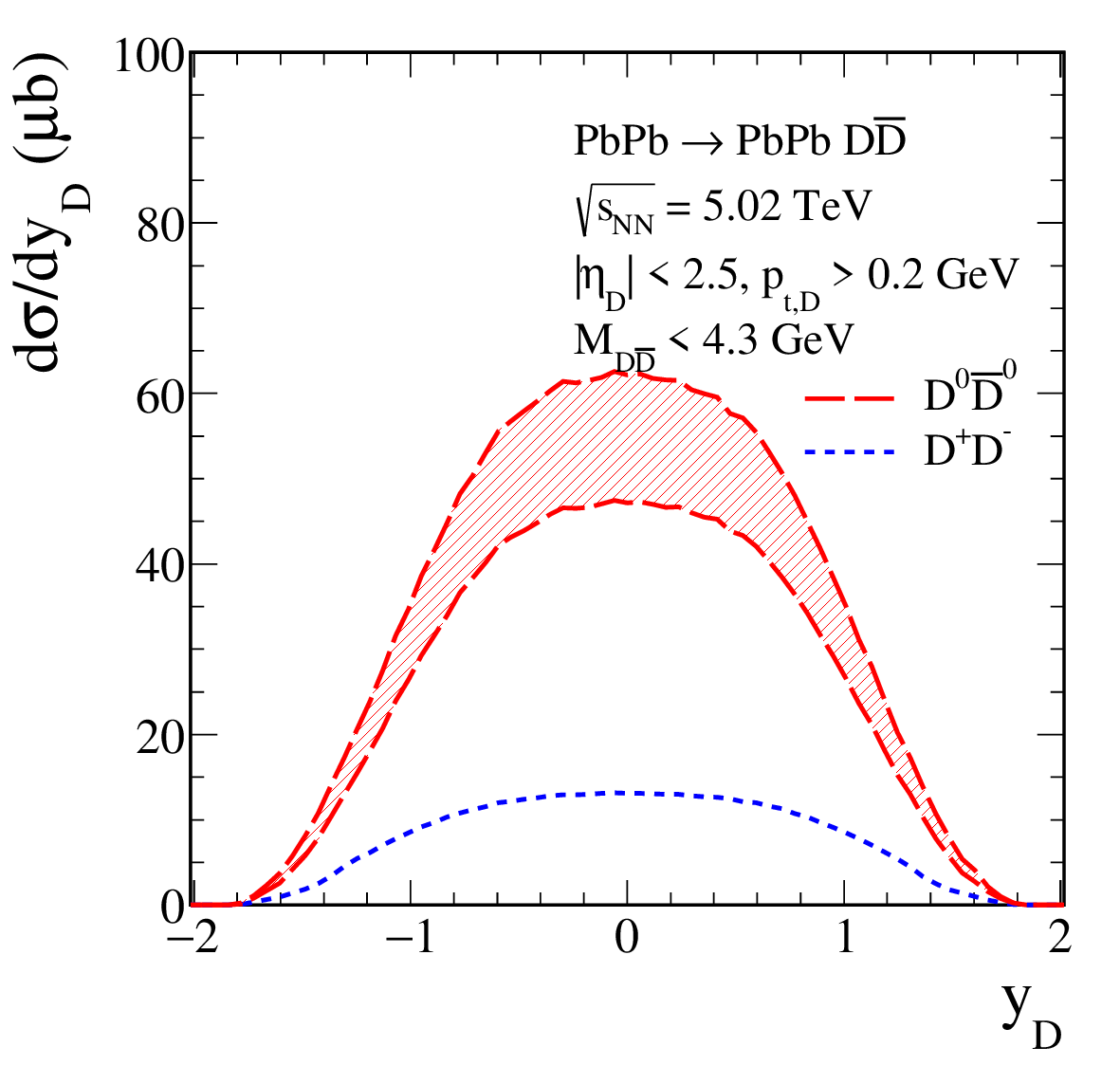}
\includegraphics[width=0.325\textwidth]{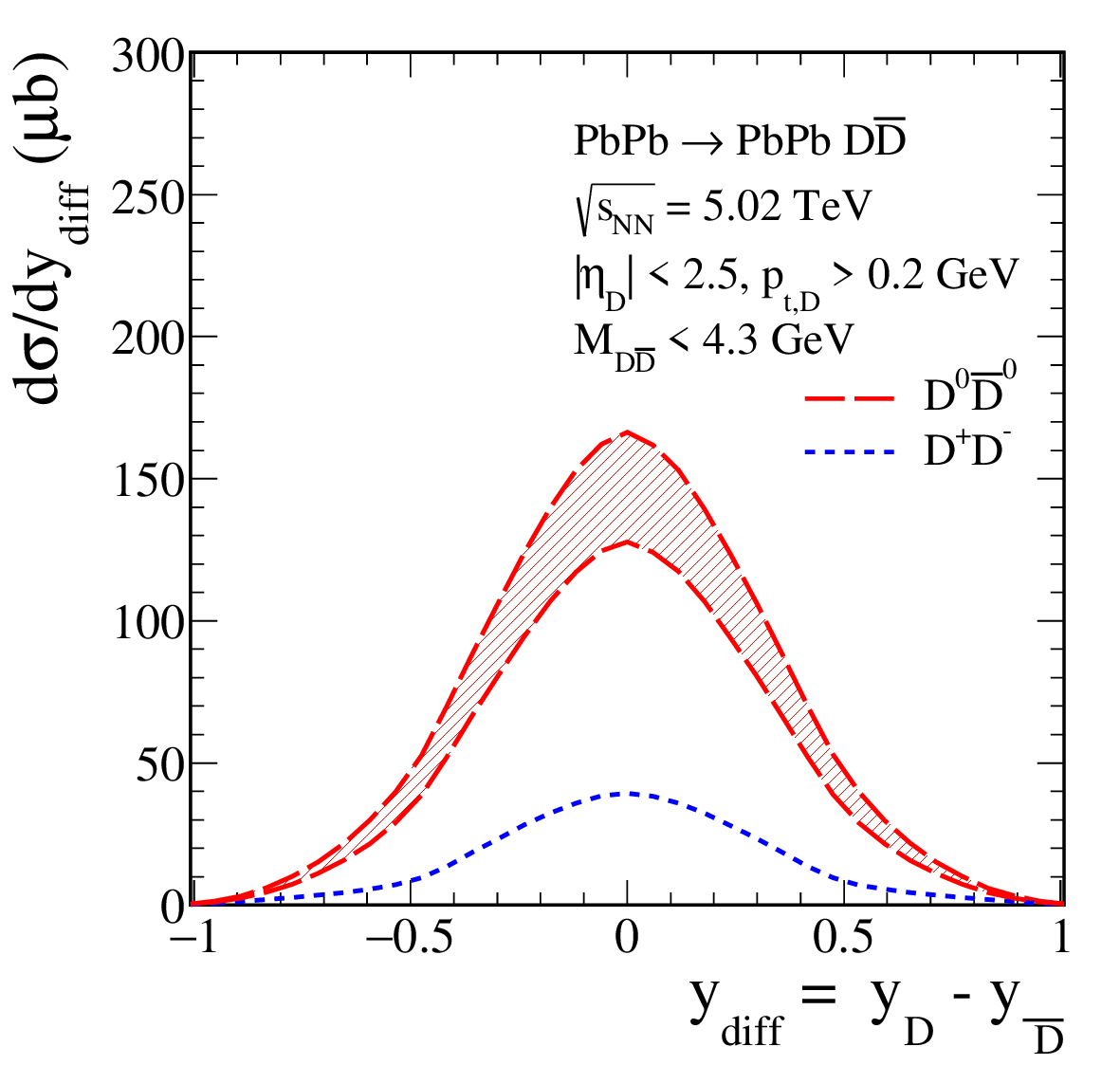}
\caption{The differential nuclear cross sections
for the reactions
${\rm Pb}{\rm Pb} \to {\rm Pb}{\rm Pb} D^{0} \bar{D}^{0}$ (the red band)
and
${\rm Pb}{\rm Pb} \to {\rm Pb}{\rm Pb} D^{+}D^{-}$
(the blue dotted curve)
calculated for $\sqrt{s_{NN}} = 5.02$~TeV and 
imposing cuts: $|\eta_{D}| < 2.5$,
$p_{t,D} > 0.2$~GeV, and $M_{D \bar{D}} < 4.3$~GeV.
For the $D^{0} \bar{D}^{0}$ channel, the continuum contribution was calculated for 
$\Lambda_{D^{*0}} = 3.3$ and 3.5~GeV in (\ref{DstarDgam_ff_exp}),
which corresponds to the lower and upper red long-dashed lines in the bands, respectively.}
\label{fig:diff_xs}
\end{figure}

Now we shall proceed to nuclear calculations.
In Fig.~\ref{fig:diff_xs} we present 
the nuclear differential cross sections
for the two reactions
${\rm Pb}{\rm Pb} \to {\rm Pb}{\rm Pb} D^{0} \bar{D}^{0}$ (see the upper curves)
and
${\rm Pb}{\rm Pb} \to {\rm Pb}{\rm Pb} D^{+}D^{-}$
(see the lower curves)
calculated at the ${\rm Pb}{\rm Pb}$ collision energy $\sqrt{s_{NN}} = 5.02$~TeV
and for the cuts specified in the figure legends.
The distributions in invariant mass of the $D \bar{D}$ pair
and transverse momenta of outgoing $D$ mesons
are particularly interesting.
We show our complete model results,
including the $D^{0} \bar{D}^{0}$ continuum,
the $D^{+} D^{-}$ continuum,
and the $\chi_{c0}(3860)$ and $\chi_{c2}(3930)$
contributions.
In the calculation, the Buchm\"uller-Tye potential
for the $\chi_{c0,2}(2P)$ states was used.
In the calculation of the $D^{+} D^{-}$-continuum contribution we neglect the off-shell meson dependences;
see the discussion after Eq.~(\ref{amplitude_DpDm_continuum}).
Our predictions indicate therefore that in such a case
we are dealing with an upper limit of the cross section
for the the $D^{+} D^{-}$-continuum contribution.
The $D^{0} \bar{D}^{0}$-continuum contribution
depends on the parameter of the off-shell form-factor
(\ref{DstarDgam_ff_exp}).
We show results for two values 
$\Lambda_{D^{*0}} = 3.3$~GeV and 3.5~GeV
corresponding to the lower and upper lines in the bands, respectively.
The dependence on $M_{D \bar{D}}$ is sensitive 
to the $p_{t,D}$ cut imposed.
The transverse momentum distributions of $D$ mesons
and $\bar{D}$ mesons are identical. 
Therefore we label them by $p_{t,D}$.
Imposing for example the cut on $p_{t,D} > 0.5$~GeV
will reduces the cross section significantly,
especially the $\chi_{c0}(3860)$ contribution.
Less suppression would be present for $\chi_{c2}(3930)$
where the maximum of the distribution occurs
at $p_{t,D} \approx 0.6$~GeV.
The minimum in the
$\eta_{D}$ pseudorapidity distribution can be understood as a kinematic effect; see the discussion in Appendix~D of \cite{Lebiedowicz:2013ika}.
The rapidity distributions of the $D$ or $\bar{D}$ meson have a maximum at ${\rm y}_{D} = 0$. 
The distributions in the difference of $D$ and $\bar{D}$ rapidities,
${\rm y}_{\rm diff} = {\rm y}_{D} - {\rm y}_{\bar{D}}$,
are narrower and peaked at ${\rm y}_{\rm diff} = 0$.

In Table~\ref{tab:1} we present the nuclear cross sections 
for the considered processes
${\rm Pb}{\rm Pb} \to {\rm Pb}{\rm Pb} 
(\gamma \gamma \to D \bar{D})$, 
for the neutral and charged final states,
calculated at $\sqrt{s_{NN}} = 5.02$~TeV and 5.36~TeV.
Various experimental cuts on $D$ and $\bar{D}$
pseudorapidities and transverse momenta
are taken into account.
The results are integrated 
in the interval $2 m_{D} < M_{D \bar{D}} < 4.3$~GeV
(as for the $e^{+} e^{-} \to e^{+} e^{-} D \bar{D}$ reaction).
The results given in Table~\ref{tab:1}
correspond to our complete model,
including the continuum and $\chi_{c0,2}(2P)$ contributions.
\begin{table}[h!]
\caption{Cross sections for the exclusive
${\rm Pb}{\rm Pb} \to {\rm Pb}{\rm Pb} D \bar{D}$ reactions
calculated for $\sqrt{s_{NN}} = 5.02$~TeV and 5.36~TeV
with the limitations on the pseudorapidities and transverse momenta of the final state $D$ mesons.
The results are obtained with the restriction $M_{D \bar{D}} < 4.3$~GeV.
The cross sections for the $D^{0} \bar{D}^{0}$ production are calculated with $\Lambda_{D^{*0}} = 3.3$ and 3.5~GeV in (\ref{DstarDgam_ff_exp}),
corresponding to the lower and upper values, respectively.}
    \centering
    \begin{tabular}{l|l|c|c|c|c}
    \hline
    \hline
\multicolumn{2}{c|} {} &
\multicolumn{2}{c|}{$\sigma ({\rm Pb}{\rm Pb} \to {\rm Pb}{\rm Pb} D^{0} \bar{D}^{0})$ ($\mu$b)} &
\multicolumn{2}{c}{$\sigma ({\rm Pb}{\rm Pb} \to {\rm Pb}{\rm Pb} D^{+} D^{-})$ ($\mu$b)} \\
     \hline
\multicolumn{2}{c|} {Cuts}     &
$\sqrt{s_{NN}} = 5.02$~TeV & $\sqrt{s_{NN}} = 5.36$~TeV &
$\sqrt{s_{NN}} = 5.02$~TeV & $\sqrt{s_{NN}} = 5.36$~TeV \\
     \hline
$|\eta_{D}| < 1.0$
& $p_{t,D} > 0.2$~GeV &
17.6 -- 23.2 & 18.3 -- 24.1&
5.5 & 5.7\\
$|\eta_{D}| < 2.5$
& $p_{t,D} > 0.2$~GeV &
99.6 -- 131.5 & 103.5 -- 136.6 &
29.3 & 30.5\\
$|\eta_{D}| < 2.5$
& $p_{t,D} > 0.5$~GeV &
62.2 -- 82.6 & 64.6 -- 85.9&
21.6 & 22.5\\
$2 < \eta_{D} < 4.5$
& $p_{t,D} > 0.2$~GeV &
59.9 -- 78.7 & 62.6 -- 82.2 &
16.4 & 17.1\\
         \hline
         \hline
    \end{tabular}
    \label{tab:1}
\end{table}

\begin{figure}
\includegraphics[width=0.325\textwidth]{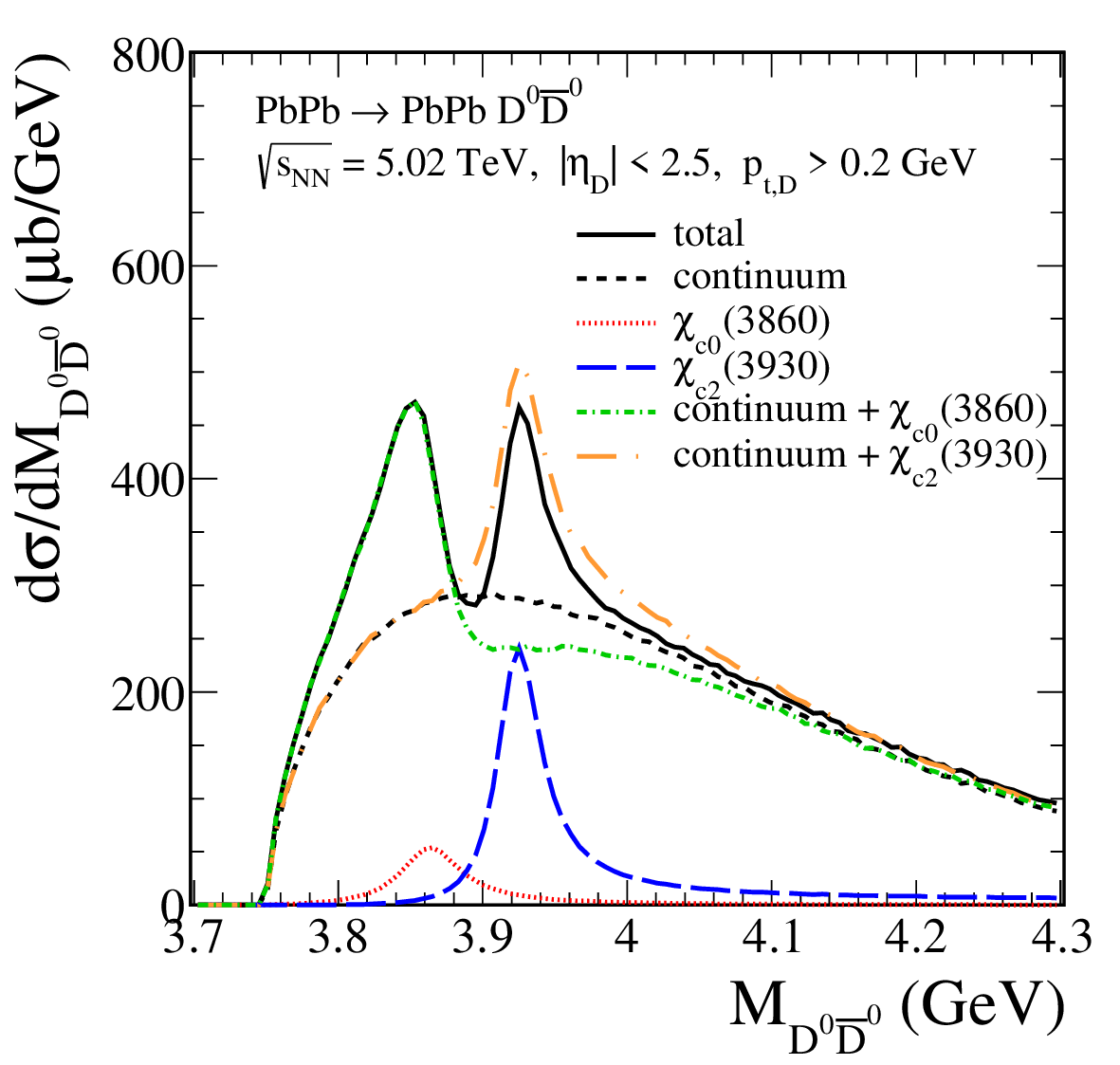}
\caption{The distribution in $D^{0} \bar{D}^{0}$ invariant mass
for the ${\rm Pb}{\rm Pb} \to {\rm Pb}{\rm Pb} D^{0} \bar{D}^{0}$ reaction.
The calculations were done for $\sqrt{s_{NN}} = 5.02$~TeV, $|\eta_{D}| < 2.5$ and $p_{t,D} > 0.2$~GeV.
The result for the continuum contribution corresponds to
$\Lambda_{D^{*0}} = 3.5$~GeV.}
\label{fig:M_deco}
\end{figure}

In Fig.~\ref{fig:M_deco} we present results for the $D^{0} \bar{D}^{0}$ invariant mass distributions
for all (total) and for individual contributions.
We show also results that correspond to the coherent sum
of both continuum and $\chi_{c0,2}$ charmonium contributions.
We wish to point out here that 
the interference of the continuum and $\chi_{c0}(3860)$ contributions 
is more significant than the one resulting from the continuum and $\chi_{c2}(3930)$.

For completeness, we give the cross sections 
for the resonant contributions
for the ${\rm Pb}{\rm Pb} \to {\rm Pb}{\rm Pb} 
(\gamma \gamma \to \chi_{c0,2} \to D \bar{D})$ reactions
calculated at $\sqrt{s_{NN}} = 5.02$~TeV.
In the calculations we use the following cuts: $|\eta_{D}| < 2.5$,
$p_{t,D} > 0.2$~GeV.
For neutral channel we obtain
\begin{eqnarray}
\sigma({\rm Pb}{\rm Pb} \to {\rm Pb}{\rm Pb} 
(\gamma \gamma \to \chi_{c0}(3860) \to D^{0} \bar{D}^{0})) &=& 3.89~\mu{\rm b}\,,  \nonumber \\
\sigma({\rm Pb}{\rm Pb} \to {\rm Pb}{\rm Pb} 
(\gamma \gamma \to \chi_{c2}(3930) \to D^{0} \bar{D}^{0})) &=& 15.24~\mu{\rm b}\,.
\label{xsec_chic0}
\end{eqnarray}
For charged channel we obtain
\begin{eqnarray}
\sigma({\rm Pb}{\rm Pb} \to {\rm Pb}{\rm Pb} 
(\gamma \gamma \to \chi_{c0}(3860) \to D^{+} D^{-})) &=& 3.64~\mu{\rm b}
\,,  \nonumber \\
\sigma({\rm Pb}{\rm Pb} \to {\rm Pb}{\rm Pb} 
(\gamma \gamma \to \chi_{c2}(3930) \to D^{+} D^{-})) &=& 13.50~\mu{\rm b}\,.
\label{xsec_chic2}
\end{eqnarray}

\section{Conclusions}
\label{sec:Conclusions}

In this Letter, we investigated the production of neutral and charged $D \bar{D}$ pairs 
in ultraperipheral ${\rm Pb}$-${\rm Pb}$ collisions (UPCs) at the LHC. 
Previous theoretical paper \cite{Luszczak:2011js} on the subject 
focussed on larger $D \bar{D}$ invariant mass region 
($M_{D \bar{D}} > 4$~GeV)
and large transverse momenta of both charmed mesons $p_{t,D} > 1$~GeV.
In our work we have tried
to incorporate the recently developed mechanisms
for the $\gamma \gamma \to D \bar{D}$ processes \cite{Babiarz:2025sld}.
The model considered here involves the non-resonant (continuum) contributions 
and $s$-channel $\chi_{c0}(3860)$ and $\chi_{c2}(3930)$ contributions
and provides a reasonable description of the
Belle and BaBar data
for $M_{D \bar{D}}$ region up to 4.3~GeV 
measured in $e^{+}e^{-}$ collisions.
The $\chi_{c0}(3860)$ and $\chi_{c2}(3930)$ mesons
are interpreted to be the first excited $P$-wave charmonia 
$\chi_{c0}(2P)$ and $\chi_{c2}(2P)$, respectively.
The $\gamma \gamma \to \chi_{c0,2}(2P)$ vertices were
evaluated within the light-front approach (NRQCD limit).
We have presented results for 
the invariant mass and angular distributions 
for the $\gamma \gamma \to D \bar{D}$
and compared to the results
obtained by the Belle and BaBar Collaborations.

We have made predictions for ultraperipheral ${\rm Pb}$-${\rm Pb}$ collisions 
at $\sqrt{s_{NN}} = 5.02$~TeV and $\sqrt{s_{NN}} = 5.36$~TeV,
and various experimental cuts for the LHC experiments,
focusing on low invariant masses of the $D \bar{D}$ system.
We have used the impact-parameter-dependent
equivalent photon approximation.
Corresponding total cross sections 
and differential distributions have been presented. 
The nuclear cross sections for the $D^{0} \bar{D}^{0}$ and $D^{+} D^{-}$ channel
calculated at $\sqrt{s_{NN}} = 5.02$~TeV
and for $|\eta_{D}| < 2.5$,
$p_{t,D} > 0.2$~GeV, and $M_{D \bar{D}} < 4.3$~GeV,
are about 115~$\mu$b and 30~$\mu$b, respectively.
The $D\bar{D}$ production in heavy-ion UPCs 
may provide new information compared 
to the presently available Belle and BaBar data, 
in particular, 
if the resonant structures of the $M_{D\bar{D}}$ distributions shown in Figs.~\ref{fig:diff_xs}
and \ref{fig:M_deco} can be observed.

In order to explore our exclusive $\gamma \gamma \to D \bar{D}$ contribution 
in UPCs one would need to go to small transverse momenta
of $D$ mesons ($p_{t,D} < 1$~GeV) and
to low $D \bar{D}$ accoplanarity 
(${\rm Aco} = 1 - \phi_{D \bar{D}}/\pi \simeq 0$,
where $\phi_{D \bar{D}}$ is azimuthal angle between the $D$ mesons), 
i.e. the back-to-back configuration between the charm mesons.
The question arises: 
Are these experimental conditions sufficient 
to separate our mechanism 
from other exclusive processes,
such as the diffractive photoproduction mechanism 
($D \bar{D}$ production via $\gamma$-Pomeron fusion processes)?
A detailed feasibility study is required to address this topic, 
which is beyond the scope of the present paper.

So far only inclusive photoproduction of single $D^{0}$ (or $\bar{D}^{0}$) mesons
in ultraperipheral lead-lead collisions at the LHC 
was measured \cite{CMS:2025tql,CMS:2025jjx} (with $p_{t,D} > 2$~GeV). 
In principle, one could also study $D^{0}$-$\bar{D}^{0}$ correlations.
The underlying mechanism is production of $c \bar{c}$ pairs 
via photon-gluon fusion associated 
by hadronization $c \to D^{0}$ and $\bar{c} \to \bar{D}^{0}$;
see e.g. \cite{Gimeno-Estivill:2025rbw,Goncalves:2025wwt,Cacciari:2025tgr}.
The mechanisms of diphoton $c \bar{c}$ pair production 
was discussed in detail in \cite{Klusek-Gawenda:2010njl}.
The corresponding nuclear cross sections for the QED mechanism
seem smaller than those found in the photon-gluon fusion processes.
The hadronization is a complicated process which is usually approximated
by the hadronization branching fraction for $c \to D$.
However, the production of $D^{0}$ or $\bar{D}^{0}$ must be associated
by production of light mesons.
Then initial back-to-back correlation between $c$ and $\bar{c}$
is destroyed and $D^{0}$ and $\bar{D}^{0}$ are produced
in a broad range of relative azimuthal angle.

We hope that the results presented here
provide a material for the study of $P$-wave higher-excited charmonia,
e.g. the $\chi_{c0}(2P)$ and $\chi_{c2}(2P)$ states,
in future measurements at the LHC.



\end{document}